# Improvement of vacuum squeezing resonant on the rubidium $D_1$ line at 795 nm

**Yashuai Han**[1,2], **Xin Wen**[1,2], **Jun He**[1,2,3], **Baodong Yang**[1,2], **Yanhua Wang**[1,2],

**and Junmin Wang**[1,2,3,]*

*1). State Key Laboratory of Quantum Optics and Quantum Optics Devices (Shanxi University),*
*2). Institute of Opto-Electronics, Shanxi University,*
*3). Collaborative Innovation Center of Extreme Optics (Shanxi University),*
*No.92 Wu Cheng Road, Tai Yuan 030006, Shan Xi Province, People's Republic of China*

\*Corresponding author, email: wwjjmm@sxu.edu.cn

**Abstract:** We report on efficient generation of second harmonic laser and single-mode vacuum squeezed light of 795 nm with periodically poled KTiOPO$_4$ (PPKTP) crystals. We achieved 111 mW of ultra-violet (UV) light at 397.5 nm from 191 mW of fundamental light with a PPKTP crystal in a doubling cavity, corresponding to a conversion efficiency of 58.1%. Using the UV light to pump an optical parametric oscillator with a PPKTP crystal, we realized -5.6 dB of a maximum squeezing. We analyzed the pump power dependence of the squeezing level and concluded that the UV light induced losses limit the improvement of the squeezing level. The generated squeezed light has huge potential application in quantum memory and ultra-precise measurement.

## 1. Introduction

The squeezed lights corresponding to the transitions of alkali metal atoms such as cesium (Cs) and rubidium (Rb) have great potential application, including nonclassical spectroscopy [1], light-atom interaction [2, 3], information storage and readout [4, 5], quantum information network [6], and ultra-precise measurement [7-11]. The performance of these applications is limited by the squeezing level. Hence it is important to generate highly squeezed light.

In recent years, the generation of squeezed lights by an optical parametric oscillator (OPO) has benefited from advances in nonlinear crystals, low-loss coatings, and high-efficiency detectors. So far, the maximum squeezing values of -12.7 dB [11] and -12.3 dB [12] have been reported from periodically poled KTiOPO$_4$ (PPKTP) crystals. Unfortunately, these results were observed at 1064 nm and 1560 nm, which are far from alkali metal atomic transitions. The great step forward was achieved by Takeno *et al.* and effective squeezing of -9 dB was observed [13] at 860 nm, which is close to Cs $D_2$ line. Afterwards, Burks *et al.* obtained more than -3 dB of squeezing for sideband frequency down to 50 kHz at 852 nm, locked on Cs $D_2$ line [14]. Furthermore, a great effort has also been dedicated to the generation of the squeezed light resonant on Rb $D_1$ line. A few of promising approaches to generating the squeezed light at this wavelength have been implemented, including the standard OPO and the atomic squeezer based on polarization self-rotation (PSF) effect and four-wave mixing. In 2002, A. B. Matsko *et al.* predicted the squeezing of -6 dB in hot vapors of $^{87}$Rb based on PSF effect [15]. After then, many experiments which generate vacuum squeezing in Rb vapors via PSF have been reported [16-19]. So far, the best squeezing via PSF in hot atomic ensemble is -3 dB [18]. Recently, the vacuum squeezed light at Rb $D_1$ line is reported via four-wave mixing and a typical result of -4 dB is achieved [20]. All of above we discussed are single-mode squeezing at 795 nm. The two-mode relative-intensity squeezed light has also been generated by four-wave mixing in Rb vapors [21, 22] and the maximum noise reduction is -9.2 dB [21]. Using OPOs, to date -2.75 dB [23] and -5.2 dB [24] of single-mode squeezed vacuum at 795 nm have been achieved, respectively. Remarkably, Predojevic *et al.* generated -2.5 dB of squeezed vacuum [25] in an OPO pumped by a diode-laser pumped system and they improved the squeezing level to -3.6 dB [26] in 2010. The generated single-mode [19, 26] and two-mode [22] squeezed lights have been applied in the magnetometer to improve the sensitivity.

Compared with the obtained squeezing at 860 nm and the longer wavelength, the generated squeezing single-mode at 795 nm via OPO is limited. However, very few papers discussed what factors limit the squeezing level at this wavelength. In this article, the maximum squeezing value of -5.6 dB at 795 nm is achieved. To our knowledge, this is the highest single-mode squeezing level at 795 nm so far. Although the improvement is not so great compared with the previous results at this wavelength, we show a detailed description on efficient generation of second harmonic laser and vacuum squeezing of 795 nm. More importantly, we analyze the power dependence of the squeezing level and the experimental data are in good agreement with the theory. We conclude that the ultra-violet (UV) light induced losses limit the squeezing level.

## 2. Experimental setup

A schematic of experimental setup is shown in Fig. 1. The fundamental source is a cw Ti:Sapphire laser pumped by a 10 W solid state pump laser at 532 nm. The wavelength of the laser is adjusted to 794.975 nm. Then we tune the laser continuously and observe the saturation absorption spectroscopy of $D_1$ transitions of Rb atoms. This ensures that the laser is quite close to the atomic transitions. A 30-dB optical isolator is utilized to avoid optical feedback. Then the beam is phase modulated at 3.6 MHz by an electro-optic modulator (EOM). This modulation is utilized to lock the enhancement cavities by using of the Pound Drever - Hall modulation sideband method [27]. The 795 nm beam is divided into four parts. The most of beam is injected into a second harmonic generation (SHG) cavity. The doubling output at 397.5 nm is used to pump the single-resonant OPO cavity. A fraction of the beam is used as a local beam for the homodyne detection after passing through a mode-cleaner (MC) cavity. The remaining two beams are used as the lock beam and the probe beam for the OPO cavity, respectively. The probe and lock beams are both injected into the cavity through a high-reflectivity coated plane mirror at 795 nm but in the counter-propagating direction. The lock beam ensures the stable generation of squeezed vacuum in the absence of the probe beam.

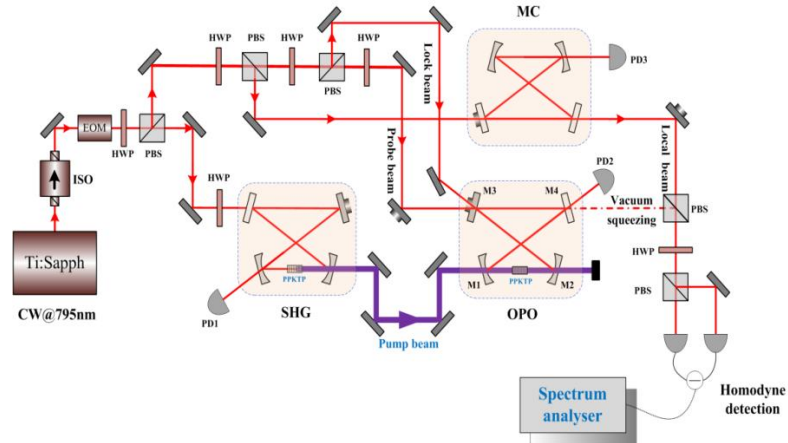

Fig. 1 Schematic of experimental setup. SHG: second harmonic generation, OPO: optical parametric oscillator, PPKTP: periodically poled KTiOPO$_4$ crystal, ISO: optical isolator, EOM: electro-optic modulator, HWP: half-wave plate, PBS: polarization beam splitter, PD: photo detector, MC: mode-cleaner cavity.

Both of the SHG and OPO cavities are designed in a uniform bow-tie configuration with a folding angle 8 °. They are made of two plane mirrors and two curved mirrors with curvature radius of 100 mm. Most of the mirrors for the enhancement cavities are super-polished in order to reduce the scattering loss as much as possible. One of the plane mirrors for the SHG cavity has a transmissivity of 10% at 795 nm, which is used as the input coupler. For the OPO cavity, the output coupler has a transmissivity of 11.5% at 795 nm. The other mirrors are all high-reflectivity coated at 795 nm. Meanwhile, the curved mirrors are also anti-reflected at 397.5 nm. Two PPKTP crystals (Raicol Crystals Ltd.) with a poling period of 3.15 μm are placed at the waist between the two curved mirrors of the SHG and OPO cavities, respectively. We choose PPKTP crystals as the nonlinear conversion materials due to a high nonlinearity together with a relatively good transmission at 397.5 nm. The crystals are placed in red copper ovens, whose temperature is stabilized by the Peltier elements and the temperature controllers. The length of both SHG and OPO cavities is 600 mm and the distance between the curved mirrors is 120 mm, yielding the beam waist of 40 μm at the central of the crystals. The MC cavity is also designed in the similar configuration in order to generate the same spatial mode as the OPO cavity. This ensures a high interference visibility in the homodyne detection, which is beneficial to achieving high level squeezing.

**3. Experimental results and discussions**

Firstly we measure the property of the SHG cavity with a 20 mm-long PPKTP crystal and the results are shown in Fig. 2. It should be emphasized that we carefully optimize the temperature of the crystal to maximize the conversion efficiency. For the mode-matched infrared power of 20 mW (240 mW), optimal phase-matching temperature is achieved for an oven temperature 53.46 °C (53.05 °C). The absorption of UV beam leads to heating of the crystal, thus the oven temperature has to be lower to compensate the temperature rise. The blue circles and red squares in Fig. 2 are experimental data. We also calculate the SHG output and the doubling efficiency versus the fundamental wave input with the following Eq. [28, 29]

$$\sqrt{\eta} = \frac{4T_1\sqrt{E_{nl}P_\omega}}{[2-\sqrt{1-T_1}(2-L_1-\Gamma\sqrt{\frac{\eta P_\omega}{E_{nl}}})]} \quad (1)$$

Where $\eta = P_{2\omega}/P_\omega$ is the doubling efficiency, $T_1$ is the transmissivity of the input coupler, $L_1$ is the intracavity linear losses referred to the pump light, $E_{nl}$ is the single-pass nonlinear conversion coefficient. $\Gamma$ refers to the losses due to second order nonlinear effects and can be written as: $\Gamma = E_{nl}+\Gamma_{abs}$, here $E_{nl}$ represents loss coefficient refer to downconverted, $\Gamma_{abs}$ formulates the absorption of SHG beam by the crystal: $P_{abs}=\Gamma_{abs}P_c^2$, where $P_c$ is the intracavity circulating power. As the SHG wavelength approaches the edge of the transparency window of KTP crystal (350 ~ 4400 nm), the absorption in the crystal is severe and can't be neglected. The absorption coefficient is about 20%/cm, as mentioned at in [30]. We adopt this value in our simulation calculation, and find that the corresponding $\Gamma_{abs}$ is $0.22E_{nl}$. Based on the other parameters of our system ($T_1$=10%, $E_{nl}$=2.3%/W, $L_1$=1.5%), we perform the numerical simulation, and the results are also shown in Fig. 2 (the blue and red lines).

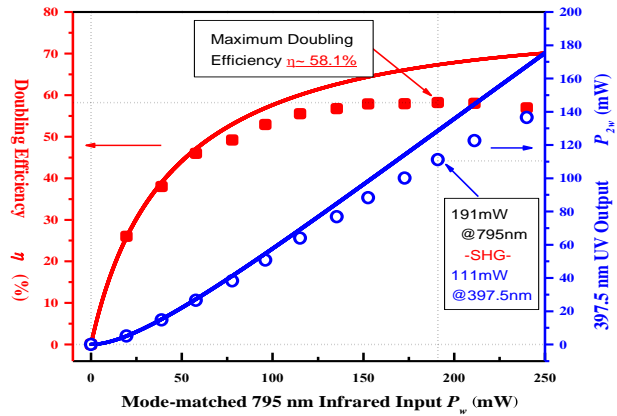

Fig. 2 The measured UV output power (blue circles) and the doubling efficiency (red squares) of the SHG cavity versus the mode-matched fundamental input. The solid lines show numerical simulation results (blue and red lines).

The measured results are in good agreement with the numerical simulation when $P_\omega \leq 150$ mW. After this value, the SHG power doesn't increase as the theoretical expectation any more. Meanwhile, the doubling efficiency goes to saturation. We achieved 111 mW of 397.5 nm UV output from 191 mW of fundamental wave input, with a maximum doubling efficiency of 58.1%. The normalized conversion efficiency is 304%/W. Recently, many efforts have been dedicated to the generation of UV laser at 397.5 nm [31-35] and blue laser at 426 nm [29, 36, 37], using PPKTP crystals placed in external cavities. Compared with the previous works, the doubling efficiency is improved. We owe this to the reduction of the intracavity losses. To our knowledge, this is the highest doubling efficiency with PPKTP at 795 nm so far. After that, the doubling efficiency begins to drop slightly as the fundamental power increases. This is due to the thermally induced bistability, which is caused by the absorption of the UV beam inside the crystal. As we discussed in our previous papers [32, 33], this thermally induced bistability prevents locking of the cavity to the top of the distorted fringe, leading to the decline of the doubling efficiency. In order to keep a stable locking of the cavity, we restrict the maximum fundamental power to 240 mW, corresponding to a UV output of 137 mW.

In order to efficiently pump the OPO cavity, the UV pump beam should be precisely mode-matched to the OPO cavity mode. The standard mode-matching method is to observe the cavity length-scanned transmission spectrum. As the cavity mirrors of the OPO are nearly transparent at 397.5 nm, no resonant build-up occurs and this method is infeasible. We use the following method. We run the OPO as a frequency doubler and the fundamental beam is coupled into the OPO cavity through the output coupling mirror M4 in the propagating direction of the squeezed light output. The generated UV beam represents the OPO cavity mode and escapes from M2. It is overlapped with the UV pump beam after passing through the OPO cavity mirror M1, PPKTP crystal, and the OPO cavity mirror M2. Scanning their relative phase, we use a photo detector to monitor their interference. With the interference visibility aligned to the maximum, we ensure that the UV pump beam is efficiently mode-matched to the OPO cavity. It is emphasized that the UV power should be as low as possible to avoid the parametric process. With this method, a maximum spatial mode-matching efficiency of 96% is achieved.

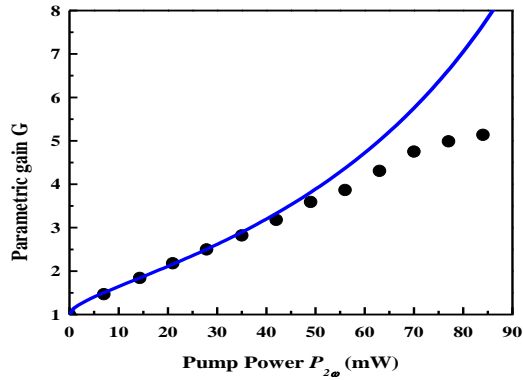

Fig. 3 Observed parametric gain versus the violet pump power.

After optimizing the mode-matching of the UV pump beam, we inject a weak signal (probe) beam into the OPO cavity through the high-reflectivity mirror M3 to run it as an optical parametric amplifier (OPA). Scanning the phase of the probe beam, we detect the phase-sensitive amplification process or the de-amplification process, caused by the nonlinear interaction with the UV pump light. At this time, the OPO cavity is locked by the counter-propagating lock beam. The measured parametric gain $G$ versus the UV pump power is shown in Fig. 3. The black circles are experimental data, where the UV pump power is actually measured in front of the OPO cavity. Similar to the SHG process, the crystal oven temperature of the OPO should also be optimized to maximize the parametric gain $G$ at each pump power. For the pump power of 15 mW (84 mW), optimal phase-matching temperature is achieved for an oven temperature of 52.75 °C (52.20 °C). The phase-matching temperature is slightly different from the 20 mm-long PPKTP crystal, due to the subtle difference in the poling period. The blue curve in Fig. 3 is the numerical simulation result according to the following Eq. [23]

$$G = \frac{1}{(1-\sqrt{P_{2\omega}/P_{th}})^2} \tag{2}$$

The threshold $P_{th}$ of the OPO cavity can be estimated with the following Eq. [23]

$$P_{th} = \frac{(T_2 + L_2)^2}{4E_{nl}^*}/\alpha \tag{3}$$

Where the total mode-matching efficiency $\alpha = 93\%$, which includes the spatial mode-matching efficiency (96%) and the transmissivity (97%) of the input coupler M1 for the UV pump light. The intracavity losses are directly measured from the finesse after replacing the output coupler M4 with a high-reflectivity mirror at 795 nm. The intracavity losses $L_2$ are calculated to be 0.4% from the measured finesse $F = 1570$. The single-pass conversion coefficient for the 15 mm-long PPKTP crystal is measured to be 1.85%/W. The transmissivity $T_2$ of the output coupler is 11.5%. Using these experimental parameters, the threshold $P_{th}$ is calculated to be 206 mW.

The blue curve shown in Fig. 3 is plotted with the Eq. (2) and the calculated threshold. It is in good agreement with the experimental results at low UV power region. With the UV power increasing, the parametric gain $G$ gradually deviates from the theoretical expectation. This phenomenon may be attributed to UV light induced losses, which is also known as the "grey tracking" effect [38]. The intracavity losses increase after illumination with the UV pump light. The intracavity losses without a pump beam are 0.4%, which is mentioned above. The losses increase up to 1.0% after the PPKTP crystal is illuminated at 84 mW of UV pump power for 60 s. We also measure the intracavity losses for several pump powers and the results can be expressed as the following Eq.

$$L_2(P_{2\omega}) = 0.00445 + 0.06767 P_{2\omega} \tag{4}$$

Where the unit of the power is W. This light-induced losses result in a rise of the threshold according to the Eq. (3). In the SHG process this phenomenon is more serious due to the higher intensity, which acts that the doubling efficiency degrades and the mode shape distorts. This mode distortion of UV pump beam will lead to a poor mode-matching of OPO cavity, hence increase the OPO's threshold, too. Inferred from the Eq. (2), this rise of the threshold, induced by the two factors together, will lead to the fact that the measured parametric gain $G$ is lower than the theoretical expectation at high UV power region.

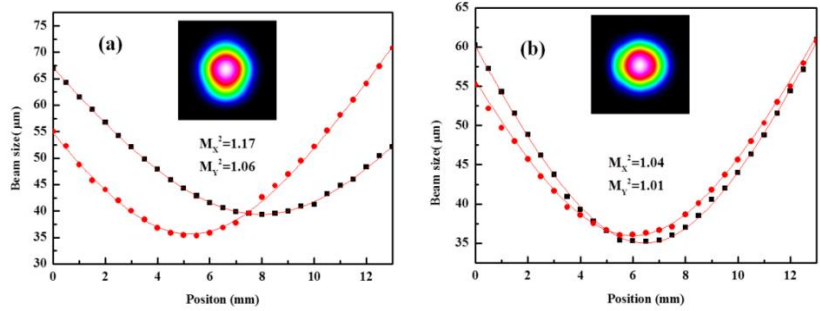

Fig. 4 The measured beam quality factors ($M^2$ values) and the beam profile of the local oscillator beam. (a) The case without the MC cavity. (b) The case with the MC cavity.

Next, we overlap the vertically polarized OPA output with the horizontally polarized local oscillator beam by using of a polarization beam splitter (PBS) cube. Passing through a half-wave plate (HWP) and another PBS, the local oscillator beam and the OPA output are mixed, balanced and collected by the home-made balanced homodyne detectors. We can detect their interference at either path of the detectors in the DC ports. The local oscillator beam is cleaned and filtered by a bow-tie MC cavity. We measure the beam profile and the beam quality factors of the Ti:Sapphire laser output, as shown in Fig. 4(a). Obvious astigmatism is observed, and the measured beam quality factors, $M_x^2 = 1.17$, and $M_y^2 = 1.06$. After the MC cavity, the astigmatism is remarkably weakened and the beam quality is significantly improved, as shown in Fig. 4(b). The measured beam quality factors, $M_x^2 = 1.04$, and $M_y^2 = 1.01$. With a help of the MC cavity, the interference visibility in the homodyne detection is improved from 96% to 99.7%.

To generate the squeezed vacuum, we block the seed beam to run the OPA as the OPO. Fig. 5 shows the measured quantum noise for the squeezed vacuum at an UV pump power of 84 mW when the local phase is scanned by using a piezo. The corresponding parametric gain $G$ is around 5.2. The noise level is measured with a spectrum analyzer in zero-span mode at an analyzing frequency of 2 MHz, with the resolution bandwidth (RBW) of 100 kHz and the video bandwidth (VBW) of 30 Hz. The maximum squeezing level of $-5.6 \pm 0.11$ dB is observed, while the anti-squeezing is $+7.0 \pm 0.11$ dB. The electronic noise is 17 dB below the shot noise level (SNL) and has been subtracted from the data.

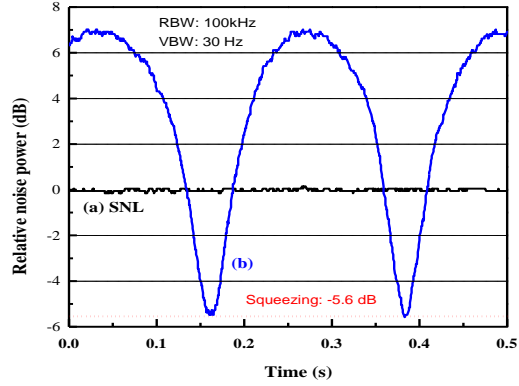

Fig. 5 Measured quantum noise for squeezed vacuum at an UV pump power of 84 mW. (a) The shot noise level (SNL) is already normalized to zero. (b) Squeezing trace when the local phase is scanned. The setting of the spectrum analyzer is zero-span mode at an analyzing frequency of 2 MHz, the resolution bandwidth (RBW): 100 kHz, the video bandwidth (VBW): 30 Hz. Trace (a) is averaged for 20 times. The electronic noise has been subtracted from the data.

We continue the experiment at several UV pump powers and summarize them in Fig. 6. The experimental data in horizontal axis are consistent with Fig. 3, so we can address the corresponding parametric gain $G$. The observed squeezing $R_-$ and anti-squeezing $R_+$ can be modeled as [12, 13]

$$R_\pm = 1 \pm \eta^* \varepsilon^2 \zeta \rho \frac{4x}{(1 \mp x)^2 + 4\Omega^2} \qquad (5)$$

Where $\eta^*$ is the quantum efficiency of the photodiode, $\varepsilon$ is the homodyne interference visibility between the squeezing output and the local oscillator beam, $\zeta$ is the propagation efficiency, $\rho = T_2/(T_2+L_2)$ is the escape efficiency of the OPO cavity, $x = (P_{2\omega}/P_{th})^{1/2} = 1-1/G^{1/2}$ is the pump parameter, $\Omega = 2\pi f/\gamma$ is the detuning parameter, $f$ is the analyzing frequency, $\gamma = c(T_2+L_2)/l$ is the decay rate of the OPO cavity, and $l$ is the total length of the OPO cavity.

In our setup, we use two photodiodes (Hamamatsu, Model S3883) with quantum efficiency of 94% in the homodyne detection system. The homodyne interference visibility $\varepsilon$ is 99.7% and the propagation efficiency $\zeta$ is 99%. The escape efficiency $\rho$ is 96.6% with $T_2 = 11.5\%$ and $L_2 = 0.4\%$. The threshold $P_{th}$ is estimated to be 206 mW. The analyzing frequency $f = 2$ MHz yields the detuning parameter $\Omega = 0.215$. Based on these parameters, we perform the numerical simulation with Eq. (5) and plot the results with black line in Fig. 6. We see that the experimental data show good agreement with theoretical results when $P_{pump} \leq 40$ mW. After this value, small discrepancies occur and the squeezing level tends to saturation. According to the theory, squeezing of -6.9 dB and anti-squeezing of +9.2 dB are expected, which are higher than the experimental observation. This discrepancy may be attributed to the unexpected parametric gain and the increase of intracavity losses with the UV pump power increasing. We take the measured gain and increased intracavity losses into account and calculate the squeezing and anti-squeezing at the fixed UV pump power, shown in Fig. 6. The measured gain is observed from Fig. 3 and the light-induced losses are modeled by Eq. (4). The measured gain $G = 5.2$ at the UV pump power of 84 mW, yields to the pump parameter $x = 0.561$. Due to the increased losses of 1.0% at this power, the escape efficiency $\rho$ decrease to 92% and the detuning parameter $\Omega$ is changed to 0.168. Using these modified parameters, the calculated squeezing and anti-squeezing are -5.7 dB and 8.0 dB, respectively. The calculated squeezing level fits well with experimental data, which confirm our analysis.

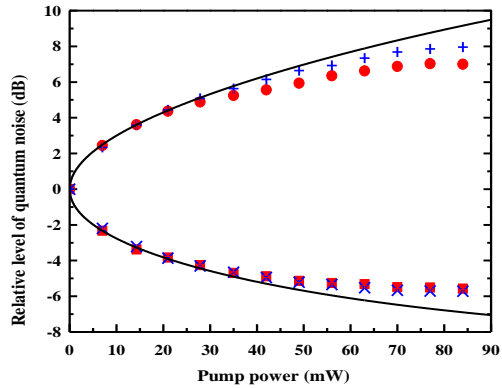

Fig. 6 Power dependence of the squeezing and anti-squeezing levels. The red dots and red squares indicate the measured results. The black lines indicate the theoretical predictions, while the blue crosses (+) and blue tilted crosses (×) indicate the calculated results using the measured parametric gain and the losses at the fixed pump power.

We operate the cavity in OPA mode to generate quadrature squeezed light and obtain the noise-power spectra at the analysis frequencies from 200 kHz to 10 MHz, shown in Fig. 7. The peak at the frequency of 3.6 MHz is caused by the modulation signal. The power of the pump field is about 42 mW. About -1.5 dB of squeezing is observed at sideband frequency down to 230 kHz. Currently, the squeezing at low sideband frequencies is mainly limited by the bandwidth of the homodyne detector.

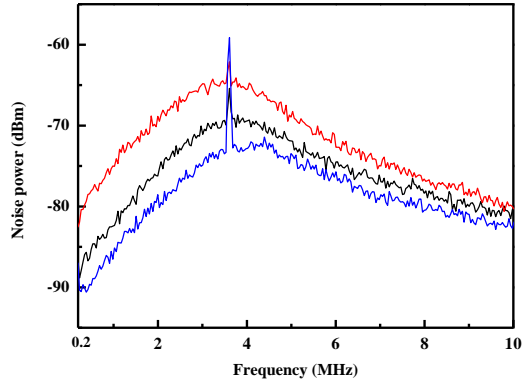

Fig. 7 Measured noise-power spectra from 200 kHz to 10 MHz. Blue line, the squeezing component; black line, the shot noise level; red line, the antisqueezing component. RBW is set to 30 kHz and VBW is 300 Hz.

## 4. Conclusion and perspective

We generate 111 mW of second harmonic laser at 397.5 nm with a conversion efficiency of 58.1% by using a PPKTP crystal. We also generate -5.6 dB of a maximum single-mode squeezing at 795 nm with the PPKTP crystal. The squeezing level is limited by the UV light induced losses. The increased losses lead to the deviation of the parametric gain from theoretical expectation and decrease the escape efficiency of OPO. As a result, the squeezing tends to saturation at high UV pump power. We just show a brief discussion on the UV light induced losses and do not have a good understanding. We guess the strong absorption of the UV pump power by the crystal induces some sort of photo refractive effect or destroys the poling period of the crystal. We hope this effect can be overcome someday, and then the squeezing at this wavelength can be improved greatly. For now, the squeezing level can also be improved to some degree by using photodiodes with higher quantum efficiency and reducing the intracavity losses. Next step, we plan to generate polarization squeezed light by controlling the relative phase between the squeezed vacuum and the local beam.

We hope the generated polarization squeezed light resonant on Rb $D_1$ line can be applied in the measurement of magnetic fields. Currently, the optical atomic magnetometer, especially the spin-exchange relaxation-free (SERF) magnetometer, is the most sensitive detector of magnetic fields. The magnetic field sensitivity has reached 0.54 fT/Hz$^{1/2}$ [39], approaching the standard quantum limit. In order to further improve the sensitivity, two limiting fundamental noise sources need to be considered, including the atomic projection noise and the optical polarization noise. The polarization squeezed probe light can reduce the optical polarization noise. Our scheme is to improve the sensitivity of Rb-based SERF magnetometer by replacing the coherent probe light with the polarization squeezed light. This system has the potential to realize a magnetometer with sensitivity beyond the state-of-the-art level. To achieve this goal, the squeezed light at a few Hz to several hundreds of kHz is required. For now, the squeezing is finally detected down to 200 kHz. In order to extend the band of squeezing down to Hz regime, several parts of noise at low sideband frequencies must be considered, including the noise of homodyne detector, the noise of probe and pump light, parasitic interference and beam jitter. Until now, we have ordered a platform with active vibration isolation and an intensity noise eater to overcome the noise of light at lower frequencies. Meanwhile, we also develop a homodyne detector, which has good performance at lower frequencies. After these efforts, we believe that the highly squeezed light at lower sideband frequencies can be successfully prepared and applied in the advanced SERF magnetometer.

## Acknowledgments

This project is supported by the National Natural Science Foundation of China (61227902, 61475091, 11274213, and 61205215), and the National Major Scientific Research Program of China (2012CB921601).

## References and Links